\documentstyle[prl,aps]{revtex}
\newcommand{\BE}{\begin{equation}}
\newcommand{\EE}{\end{equation}}
\newcommand{\BA}{\begin{eqnarray}}
\newcommand{\EA}{\end{eqnarray}}
\begin{document}
\draft
\twocolumn[\hsize\textwidth\columnwidth\hsize\csname@twocolumnfalse\endcsname

\title {Solution of the two identical ion Penning trap final state}

\author{W. Blackburn} 
\address{3633 Iron Lace Drive, Lexington, KY 40509}

\author{T. L. Brown} 
\address{Department of Electrical Engineering, Washington University, 
P.O. Box 1127, St. Louis, MO 63130}


\author{E. Cozzo}
\address{201 Ellipse Street, \#12, Berea, Kentucky 40403}

\author{B. Moyers}
\address{Wine.com, Inc.,  
665 3rd Street Suite 117
San Francisco, CA 94107}

\author{M. Crescimanno}
\address{Center for Photon Induced Processes, 
Department of Physics and Astronomy, 
Youngstown State University, Youngstown, OH, 44555-2001}
\medskip

\date{\today} \maketitle

\begin{abstract}
We have derived a closed form analytic expression for the 
asymptotic motion of a pair of identical 
ions in a high precision Penning trap. 
The analytic solution includes the effects
of special relativity and
the Coulomb interaction between the ions. 
The existence and physical relevance of such a final 
state is supported by a 
confluence of theoretical, experimental and numerical
evidence. 
\end{abstract}

]

\noindent High precision Penning traps are 
ideal for studying physical characteristics
of individual ions. These traps, as described 
for example in Ref.[1], have 
magnetic fields that over the trajectories of the ions  
vary by less than a part per billion. In consequence, 
the motional frequency linewidths can be made so narrow that
effects of special relativity are readily apparent even 
at these relatively low velocities$^2$. 

To remove systematic effects it is often 
desirable to fill the trap with 
two ions and much is known about the resulting 
frequency perturbations caused by the Coulomb interaction 
between dissimilar ions$^{3}$. 
The situation with two identical ions has also been
extensively 
studied much (see Ref.[4,5] and references therein).
The solution and approach that we describe here are
rather different than those references however, since 
they include the electric trap field but ignore 
relativistic mass increase. 
Including this effect of special 
relativity may be  crucial for understanding the 
observation$^{6}$ of cyclotron mode-locking between identical ions 
(see also Ref.[7]). 

We present details of an analytical model of 
two identical ions in a high precision Penning 
trap. The model is asymptotically solvable in 
terms of elliptic functions. 
This solution is, in practical terms for protons and 
heavier ions, a generic final state 
of two identical ions in a precision Penning trap. 

We begin with a symmetry argument detailing what is
special about the two identical ion system and then we introduce and 
solve the model. For two dissimilar ions
the center of charge is different than the center of mass. 
The motion of the center of charge causes currents to run 
in the detection circuit and in the walls of the trap itself
causing a force to act back on the ions. This retarding force 
acts on the center of charge and so if the center of 
charge is different than the center of mass these damping forces
act always on a mixture of the center of mass motion and the 
relative motions of the ion pair. 

This is not the case for identical ions in the trap. In that case
the center of mass and the center of charge are the same and so the 
retarding force acts only on the center of mass motion. Thus, the relative 
motion of the ions is relatively undamped, being subject only to 
the weaker quadrupolar damping (which is associated with timescales
generally longer than typical experiments). In this sense we speak of this
final state of the two identical ion system as a decoupled, or, dark state. 

One way to understand the existence of this cyclotron dark state is with 
a symmetry argument. Neglect dissipation, relativity and interaction
and consider the Poisson 
algebra of two ions moving in a horizontal plane 
(we shall describe why this is relevant to experiment later)
in a uniform perpendicular magnetic field. The Hamiltonian is 
proportional to 
${\cal H}=p_1^2+p_2^2 + \alpha (p_3^2+p_4^2)$, where $\alpha = m_a/m_b$ is the
mass ratio and $p_{1,2}$ (resp. $p_{3,4}$) are the canonical momenta 
of particle $a$ (resp. particle $b$). For $\alpha \neq 1$ the subalgebra 
commuting with ${\cal H}$ is $so(2)\ {\rm x}\ so(2)$ whereas 
if $\alpha=1$ the algebra is $so(2)\ {\rm x}\ so(3)$. The fact that 
there are additional commuting generators in the equal mass case indicates
that there is a flat direction in the dynamics of that case, 
corresponding to degeneracy between cyclotron dark states of different 
total angular momentum. 

There is a straightforward  geometrical way of understanding the special 
qualities of the two identical ion Penning trap. 
Again consider the ions confined to a
plane perpendicular to the magnetic field and ignore 
temporarily the effects of relativity and interaction. 
The total angular momentum of the
two  
ion system is $L = p_1^2+p_2^2 + p_3^2+p_4^2$ (note independent of the mass ratio 
$\alpha$). Now, turning on relativity and interactions perturbatively, 
we learn that the motion is
essentially restricted to the intersection of iso- ${\cal H}$ and 
iso- $L$ surfaces. A generic intersection of these surfaces in ${\bf R}^4$ 
for the 
$\alpha \neq 1$ case is a two-dimensional torus 
(and so has an isometry group 
$so(2)$ x $so(2)$) whereas when  $\alpha = 1$ the 
intersection is not generic, but is the whole $S^3$. 
Although the isometry group of $S^3$, being $so(4)$, is isomorphic to 
$so(3)$ x $so(3)$ the physically relevant isometry group is that which 
preserves not only the geometry but also the underlying Poisson structure, 
which is $sp(4)$ in this case. The canonical intersection$^8$ in 
the group of matrices $GL(4)$ of $so(4)$ and $sp(4)$ 
is the algebra $u(2)$, which is isomorphic to $so(2)$ x $so(3)$, which 
again is the enhanced symmetry discussed above. 
We note that both the geometrical and 
algebraic picture can be easily generalized to the case of 
$N$ identical ions$^9$. 

Having described the symmetry properties unique to two identical 
ions in a Penning trap, we
now introduce the interacting model by starting with the 
following three assumptions. 
\medskip

\noindent 1) The ions are very near the center of the trap,
and ignore
effects due to the spatial gradient of the electrostatic 
fields of the trap (that is, we 
completely ignore the trap magnetron motion). 
The cyclotron frequency shifts 
in an isolated ion's cyclotron motion is 
entirely due to relativistic effects.  
\medskip

\noindent 2) the ions are mode locked already in the trap's
axial drive and so their motions may be 
thought of as being confined to a plane$^{6,7}$. 
\medskip

\noindent 3) The energy loss mechanism is 
entirely due to the dissipation 
of image charge currents induced in the trap/detection system, 
and thus couple only to the center 
of mass of the ion pair). 

Under these assumptions, the equations of motion for the 
ion pair are the formidable looking non-linear coupled 
differential equations;
\BE \ddot{\vec r}_1 + \omega_0(1-f_1){\hat z}\times{\dot{\vec r}_1}
+ \gamma {\dot{\vec X}}_{cm} - {{e^2 {\hat R}}\over{m_0 R^2}} = 0
\label{4.1}
\EE
\BE \ddot{\vec r}_2 + \omega_0(1-f_2){\hat z}\times{\dot{\vec r}_2}
+ \gamma {\dot{\vec X}}_{cm} + {{e^2 {\hat R}}\over{m_0 R^2}} = 0
\label{4.2}
\EE
where ${\vec X}_{cm} = ({\vec r}_1 + {\vec r}_2)/2$, and 
${\vec R}= {\vec r}_1 -{\vec r}_2$ and where
$f_1= {{|{\dot {\vec r}}_1|^2}\over{2c^2}}$ is just the ratio of the 
kinetic energy to the rest mass-energy of ion 1
(similar expression for $f_2$ is in terms of the kinetic energy of  
the second particle). 
This term, due entirely to special relativistic mass increase, causes the 
cyclotron frequency to depend on the kinetic energy of the ion(s).

We add and subtract Eq.~(\ref{4.1}) and Eq.~(\ref{4.2}) to rewrite 
them in terms of the center of mass 
co-ordinate ${\vec X}_{cm}$ and the relative 
coordinate ${\vec R}$, 
$$ 
{\ddot {\vec X}}_{cm} + 2\gamma{\dot {\vec X}}_{cm} 
+ \omega_0(1-{{f_1+f_2}\over{2}}){\hat z}\times{\dot {\vec X}}_{cm}
$$
\BE 
= {{\omega}\over{4}}(f_1-f_2){\hat z}\times{\dot {\vec R}}
\label{4.3}
\EE

\BE 
{\ddot {\vec R}}+ \omega_0(1-{{f_1+f_2}\over{2}}) 
{\hat z}\times{\dot {\vec R}} - {{2e^2 {\vec R}}\over{m_0 R^3}}
={\omega_0} (f_1-f_2) {\dot {\vec X}}_{cm}
\label{4.4}
\EE
Let ${\vec V} = {\dot {\vec X}}_{cm}$ be a symbol for the 
center of mass velocity. As expected, 
only the center of mass {\it velocity} 
enters into the equations. Confined as they are to the same vertical plane, 
this becomes a six-dimensional (phase-space) system. 
Let ${\vec U} = {\dot {\vec R}}$. In these variables, the 
combinations $f_1-f_2= {{ {\vec U}\cdot {\vec V}}\over{c^2}}$ 
and $f_1+f_2 = {{ {\vec U}^2 + 4{\vec V}^2}\over{4c^2}}$. 

As per earlier discussion, from Eq.~(\ref{4.3}) 
and Eq.~(\ref{4.4}), it is clear that the center of mass motion 
is damped but the relative motion is not.
Thus, after sufficient
time, it is consistent to assume that the center of mass motion 
damps out completely, that is, ${\vec V} \rightarrow 0$. 
The coupling term between the ${\vec R}$ motion
and the ${\vec V}$ (center of mass) motion is through 
the term proportional to $f_1-f_2$ (itself proportional to $V$), 
and so Eq.~(\ref{4.3}) and 
Eq.~(\ref{4.4}) quickly decouple as ${\vec V}\rightarrow 0$. 

The resulting motion can be treated perturbatively in small 
${\vec V}$. To find the zeroth order term we ignore the 
coupling term completely, resulting in exponential decay for 
${\vec V}$ 
and the total center-of-mass kinetic energy. 
Asymptotically for
the relative co-ordinate 
Eq.~(\ref{4.4}) becomes
\BE  {\ddot {\vec R}}+ \omega_0(1-{{f_1+f_2}\over{2}}) 
{\hat z}\times{\dot {\vec R}} - {{2e^2 {\vec R}}\over{m_0 R^3}}
= 0 \ \ \ .
\label{4.5}
\EE
This is a system of two coupled non-linear second order differential 
equations. Generally such systems do not admit 
closed-form, analytical solution. Somewhat surprisingly, we now 
point out that Eq.~(\ref{4.5}) admits a general solution
in terms of elliptic functions. 

The approach is standard. 
First we find two integrals of the motion, reducing the four (phase space) 
dimensional system in Eq.~(\ref{4.5}) to a two dimensional (phase space) 
system. 
The integrals are the energy and 
a generalization of angular momentum.
The inter-ion energy results from  
taking the dot product of Eq.~(\ref{4.5}) with ${\dot {\vec R}}$, forming the 
total differential, and integrating to find the integration 
constant, 
\BE u_0 = {{1}\over{2}} | {\dot {\vec R}}|^2 + {{2e^2}\over{m_0 R}} 
\label{4.6}
\EE 
Since the equations have manifest rotational symmetry,
there is a conserved angular momentum. 
As always with a magnetic field, the 
total angular momentum receives a contribution from the 
magnetic field. 
Proceed by taking the vector cross 
product of ${\vec R}$ and Eq.~(\ref{4.5}) to find 
\BE {{ {\rm d}L }\over{ {\rm d}t}}
-{{\omega_0 }\over{2}}(1-f) {{{\rm d}R^2}\over{ {\rm d}t}}
=0
\label{4.7}
\EE  
where, as always, $R = |{\vec R}|$, and 
$f=({{|{\dot {\vec R}}|}\over{2c}})^2$ is the term 
due to special relativity. The angular momentum 
per unit mass $L = {\hat z}\cdot({\vec R} \times {\dot {\vec R}})
=R^2 {{ {\rm d}\phi}\over{ {\rm d}t}}$ is the standard definition.

Now, using the inter-ion energy integral Eq.~(\ref{4.6}), 
$f$ can be written entirely as a function of $R$.
Doing so for $f$ in Eq.~(\ref{4.7}) and integrating leads to 
the integration constant $L_0$, 
\BE L_0 = L-{{w_0}\over{2}}\bigl(1- {{u_0}\over{2c^2}}\bigr)R^2
-{{\omega_0e^2}\over{2m_0c^2}}R
\label{4.8}
\EE
$L_0$ represents the 
generalized angular momentum. 

Since they are independent, the constants of motion in 
equations Eq.~(\ref{4.6}) and Eq.~(\ref{4.8}) constrain the 
motion to lie in a two-dimensional surface 
in the original four-dimensional phase space. 
Of course, that fact by itself is insufficient to guarantee
integrability of the equations of motion in closed form.
However additional peculiarities of this 
system Eq.~(\ref{4.5}) result in closed form solution. 

In polar co-ordinates the kinetic energy in the potential 
energy equation can be written 
\BE \biggl({{ {\rm d}{\vec R}\over{ {\rm d}t}}}\biggl)^2 
= \biggl({{ {\rm d}R}\over{ {\rm d}t}}\biggl)^2 + 
{{L^2}\over{R^2}}
\label{4.9}
\EE
and solving Eq.~(\ref{4.8}) for $L$ and substituting 
we find that Eq.~(\ref{4.6}) becomes, 
\BE \biggl({{ {\rm d}R}\over{ {\rm d}t}}\biggl)^2 = 
2u_0-{{4e^2}\over{m_0R}}-{{ (L_0+\alpha R+\beta R^2)^2}\over{R^2}}
\label{4.10}
\EE
where $\alpha={{\omega_0e^2}\over{2m_0c^2}}$
and $\beta={{\omega_0}\over{2}}
\bigl(1-{{u_0}\over{2c^2}}\bigr)$. Since the RHS involves 
only five {\it consecutive} powers of $R$
(namely, $R^2, R, R^0, ... R^{-2}$). 
the equation is that of an elliptic function. 

More explicitly, we now compute the orbital period
of the dark state and find the orbit trajectory parametricaly. 
To compute the period we rewrite Eq.~(\ref{4.10}) as
\BE 
{\rm d}t = {{ {\rm d}R}\over{\sqrt{{\tilde u}
-L_0^2/R^2-n/R-2\alpha\beta R-\beta^2 R^2}}}
\label{4.11}
\EE
with ${\tilde u} = 2u_0-2L_0\beta-\alpha^2$, 
and $n=2L_0\alpha+{{4e^2}\over{m_0}}$. 

The integral is a combination of standard elliptic functions.
In lab co-ordinates $R,\phi$ the orbits will in general be open
(with some precession rate which can be written in terms of 
complete elliptic integrals) just as viewing the orbits 
in the $R,t$ co-ordinates, where now ``precession'' in 
$t$ in simply the period of the orbit. The period $T$ of 
these orbits is thus given by a contour integral
of the RHS of Eq.~(\ref{4.11}) around the cut running 
between the classical turning points (we label) 
$a_0$ and 
$a_1$, namely, 
$$ 
T = \int {\rm d}t = \oint {{ {\rm d}R}\over{\sqrt{{\tilde u}
-L_0^2/R^2-n/R-2\alpha\beta R-\beta^2 R^2}}}
$$
\BE
= {{1}\over{i\beta}} \oint {{ R{\rm d}R}\over{\sqrt{
(R-a_0)(R-a_1)(R-a_2)(R-a_3)}}}
\label{4.12}
\EE
where the $a_i$ are the roots of the fourth degree polynomial
written in Eq.~(\ref{4.17}). 
By looking at the signs of terms in the polynomial
we can see that there can be at most two real positive roots. 
Physically we expect there to be exactly two real positive
roots which we have called $a_0$ and $a_1$. 
These are the classical turning points of the motion, and 
represent the furthest and nearest approaches of the particles. 

Furthermore, 
in the system we are working with, for typical  
values of parameters, we find that all roots are real, with 
two positive and two negative. We may then order the 
roots $a_0 \geq a_1 \geq a_2 \geq a_3$. 
Note also that the canonical choice of phase for the 
square root on the cut  between $a_0$ and $a_1$ is 
$i$ and so the period in Eq.~(\ref{4.12}) is real and positive. 

Finally, computing the integral in Eq.~(\ref{4.12}) yields
(notation is that in Ref.\cite{10}), 
\BE T = {{2}\over{\rho \beta}} \biggl[
(a_0-a_3) \Pi( {{a_1-a_0}\over{a_1-a_2}},k)
+a_3 K(k)\biggr]
\label{4.13}
\EE
where $K$ and $\Pi$ are respectively the complete elliptic integrals
of the first and third kind, and $\rho = \sqrt{(a_0-a_2)(a_1-a_3)}$
and $k = {{\sqrt{(a_0-a_1)(a_2-a_3)}}\over{\rho}}$ 
is the square root of the cross-ratio of the roots. 
Note that the first argument in the $\Pi$ is negative, 
as it should be on physical grounds, since $\Pi$ is convergent
for any negative argument. 

One of the most striking experimental surprises of the two identical ion 
system is the discovery of cyclotron mode-locking$^{6}$. In these events the 
two frequency traces corresponding (approximately) to the individual 
ions motions meld into one trace. This visible trace is the center of 
mass motion of the dark state. Our analysis indicates that there 
is another invisible (as a dipole) frequency branch associated
with the inter-ion motion and that it has frequency ${{2\pi}\over{T}}$ with 
$T$ of Eq.~(\ref{4.13}). 
For the case of two protons in a typical 
precision Penning trap (at $\omega_0 \sim 5 {\rm x} 10^8$) 
we find that Eq.~(\ref{4.13})
yields frequencies are some tens 
of Hertz different than $\omega_0$.  
It would be an interesting test to apply 
a sequence of dipole and quadrupolar fields to make transitions 
between dark states and (visible) center of mass states. 

By standard means 
we now derive explicit formulae for the shape of
the dark state orbits. 
Recall that, by definition of the angular momentum, $L$,  
and Eq.~(\ref{4.8})
\BE 
{{ {\rm d}\phi}\over{ {\rm d}t}} = 
{{L_0}\over{R^2}}+{{\alpha}\over{R}} + \beta
\label{4.14}
\EE
Thus, eliminating time between this and Eq.~(\ref{4.11}) we 
find 
\BE 
\phi = {{1}\over{i\beta}} \int {{ R({{L_0}\over{R^2}}+
{{\alpha}\over{R}} +\beta)\  {\rm d}t}\over{
\sqrt{(R-a_0)(R-a_1)(R-a_2)(R-a_3)}}}
\label{4.15}
\EE
which may be evaluated in terms of incomplete elliptic 
functions. We find
$$
\phi-\phi_0 = 
{{2}\over{\rho}} \biggl[\bigl({{L_0}\over{a_2}}+\alpha+\beta a_2
\bigr)F(\theta(R),k)
$$
$$ 
+\bigl({{a_2}\over{a_1}}-1\bigr)
\bigl\{ {{L_0}\over{a_2}}\Pi\bigl(\theta(R), 
{{a_2(a_0-a_1)}\over{a_1(a_0-a_2)}}, k\bigr)
$$
\BE
-\beta a_2 \Pi\bigl(\theta(R), {{a_0-a_1}\over{a_0-a_2}}, k\bigr)
\bigl\}\biggr]
\label{4.16}
\EE
where, again, the $a_i$ are the (ordered) roots of the polynomial 
$$ 
P(R) = -\beta^2 R^4 -2\alpha\beta R^3 + 
(2u_0-\alpha^2-2L_0\beta)R^2  
$$
\BE
-(2L_0\alpha+{{4e^2}\over{m_0}})R
-L_0^2
\label{4.17}
\EE
with 
$\alpha$
and $\beta$
as defined previously 
and where 
\BE
\sin \theta(R) = \sqrt{
{{(a_0-a_2)(R-a_1)}\over{(a_0-a_1)(R-a_2)}}}
\label{4.18}
\EE 

Note directly from Eq.~(\ref{4.16}) and Eq.~(\ref{4.18}) 
that the precession of these orbits is
given by twice the RHS Eq.~(\ref{4.16}) with each incomplete
elliptic functions replaced by its complete 
elliptic counterpart. 

We have completed a numerical simulation of the system 
Eq.~(\ref{4.1}) and Eq.~(\ref{4.2}) 
for a range of initial conditions. To 
abet numerical stability those equations were rewritten
in the co-rotating frame and integrated using commercial (IDL$^{tm}$) 
routines on a DEC Alpha workstation. 
Some of these IDL$^{tm}$ programs 
link compiled versions of 
CERN's Mathlib elliptic function routines. The 
results from a typical run are shown in Figures 1 (resp. 2) 
where both the $u_0$ of Eq.~(\ref{4.6}) (resp. $L_0$ of Eq.~(\ref{4.8})) 
are plotted as functions of time. 

The figures show that initially the motions of the ions are 
essentially independent as the energy dissipates. During this regime 
the total energy of the system is split between the center of mass 
motion and the inter-ion motion. Note that due to the large dynamic 
range of these simulations we have plotted the logarithm of the energy. 
Thus, the linear decay of the envelope of the inter-ion energy $u_0$ 
in this initial regime is the exponential damping of the energy of the 
system as a whole.  

Eventually the center of charge motion damps away appreciably
and the remaining
inter-ion motion persists. 
As described earlier, in real 
experiments of this type the dark state we are describing 
is likely to be 
effectively the final state since we expect the inter-ion motion to decay
via quadrupole radiation on a timescale long compared with typical 
two-ion experiments. 
For our simulation this final state is reached 
at simulated time 150, after 
which both $u_0$ and $L_0$ are essentially constant
(up to numerical accuracy of the simulations). 

In conclusion, we have derived  
closed form analytic formulae for the 
dark state of two identical ions in a Penning trap. 
To find this solution, we assumed that the pair is 
near the center of the trap (we have completely neglected the 
effect of the trap's electrostatic fields)
and that the motion of the ions is confined to
the same azimuthal  plane. 
It is straightforward to include in this analysis
the effects of the trap's 
electric field and also a fixed average vertical offset 
between the cyclotron 
planes of the ions. This results in formulae for
the two integrals of motion that have additional terms compared with the 
Eq.~(\ref{4.6}) and Eq.~(\ref{4.8}). 
However, the resulting equations of motion 
for the dark state are no longer solvable in terms of known 
functions. 

This research was supported
in part by Research Corporation
Cottrell Science Award \#CC3943 and \#CC5285 in part by the National Science
Foundation under grants PHY 94-07194  and EPS-9874764
and in part by Appalachian Colleges Association 
Mellon Foundation Student-Faculty Grants. 
We would like to thank CERN Mathlib for the 
use of the elliptic function libraries.
We are delighted to thankfully acknowledge
G. Gabrielse, C. H. Tseng, D. Phillips,  
L. J. Lapidus, A. Khabbaz and A. Shapere
for many interesting and
stimulating discussions and the theory 
group at the University of Kentucky where
much of this work was done.


\newpage

\twocolumn[\hsize\textwidth\columnwidth\hsize\csname@twocolumnfalse\endcsname

\input epsf.sty

\begin{figure}
\begin{center}
\epsfxsize=5in\epsffile{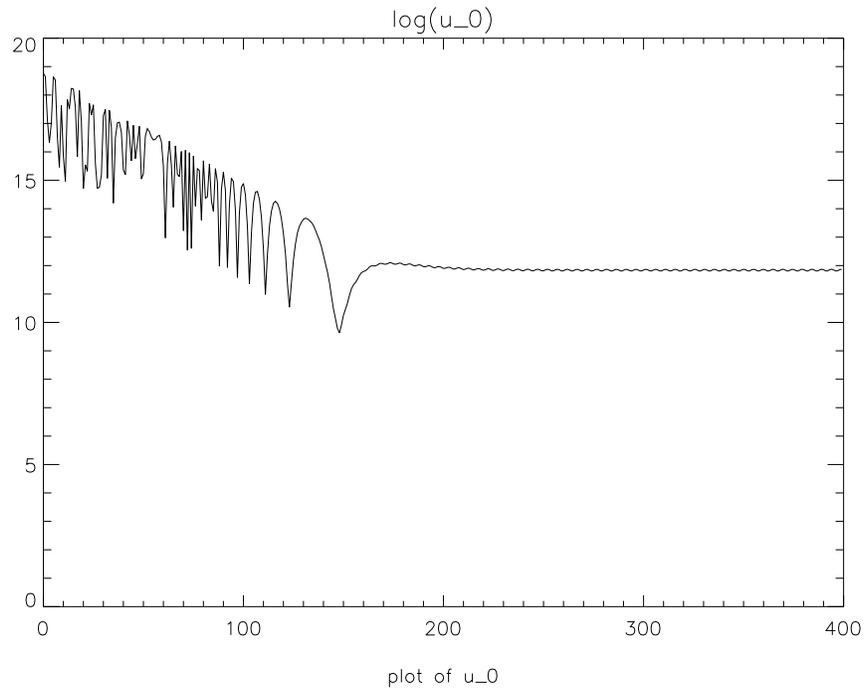} \caption{The internal energy $u_0$ as a function of time.} 
\label{fig1}
\end{center}
\end{figure}



\begin{figure}
\begin{center}
\epsfxsize=5in\epsffile{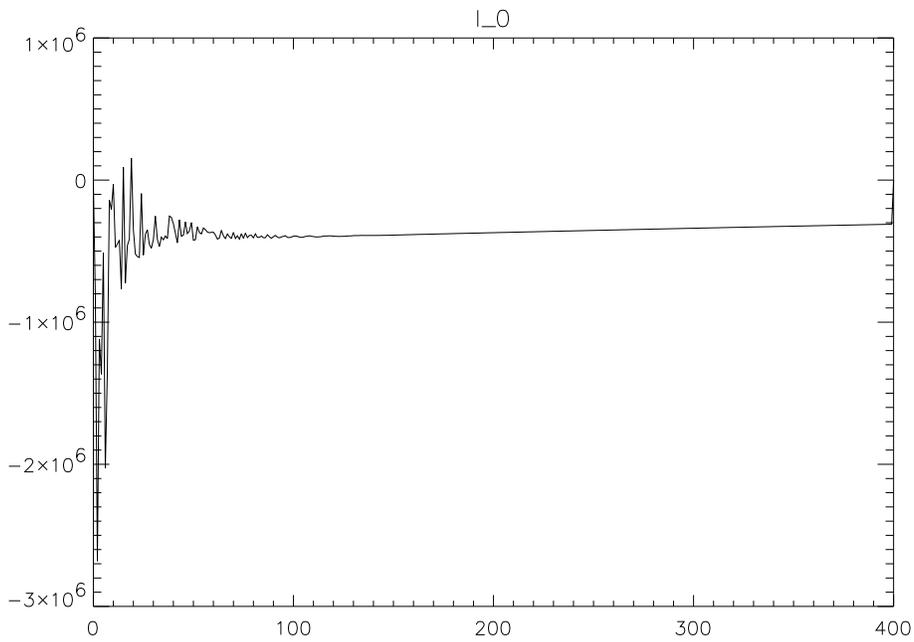} \caption{The internal angular momentum $L_0$ as a function of time.} 
\label{fig2}
\end{center}
\end{figure}
]

\end{document}